# Observation of sixfold degenerate fermions in PdSb$_2$

Xiàn Yáng,[1,\*,†] Tyler A. Cochran,[1,\*] Ramakanta Chapai,[2,\*] Damien Tristant,[2,3,\*] Jia-Xin Yin,[1] Ilya Belopolski,[1] Zǐjiā Chéng,[1] Daniel Multer,[1] Songtian S. Zhang,[1] Nana Shumiya,[1] Maksim Litskevich,[1] Yuxiao Jiang,[1] Guoqing Chang,[1] Qi Zhang,[1] Ilya Vekhter,[2] William A. Shelton,[3] Rongying Jin,[2] Su-Yang Xu,[1] and M. Zahid Hasan[1,4,5,†]

[1]*Laboratory for Topological Quantum Matter and Advanced Spectroscopy (B7), Department of Physics, Princeton University, Princeton, New Jersey 08544, USA*
[2]*Department of Physics and Astronomy, Louisiana State University, Baton Rouge, Louisiana 70803, USA*
[3]*Cain Department of Chemical Engineering, Louisiana State University, Baton Rouge, Louisiana 70803, USA*
[4]*Princeton Institute for Science and Technology of Materials, Princeton University, Princeton, New Jersey 08544, USA*
[5]*Materials Sciences Division, Lawrence Berkeley National Laboratory, Berkeley, California 94720, USA*

Three types of fermions have been extensively studied in topological quantum materials: Dirac, Weyl, and Majorana fermions. Beyond the fundamental fermions in high energy physics, exotic fermions are allowed in condensed matter systems residing in three-, six- or eightfold degenerate band crossings. Here, we use angle-resolved photoemission spectroscopy to directly visualize three-doubly-degenerate bands in PdSb$_2$. The ultrahigh energy resolution we are able to achieve allows for the confirmation of all the sixfold degenerate bands at the R point, in remarkable consistency with first-principles calculations. Moreover, we find that this sixfold degenerate crossing has quadratic dispersion as predicted by theory. Finally, we compare sixfold degenerate fermions with previously confirmed fermions to demonstrate the importance of this work: our study indicates a topological fermion beyond the constraints of high energy physics.

Condensed matter systems host exotic quasiparticles that are attracting ever increasing interest. They provide a rich platform for studying fermions inaccessible in high energy physics. For example, three-dimensional Dirac [1-5] and Weyl fermions [6-11] in solids were visualized in solid state quantum materials. Dirac fermions arise in nonmagnetic, centrosymmetric materials when two spin-degenerate bands linearly cross at a fourfold degeneracy. If time-reversal or inversion symmetry is broken, the fourfold degeneracy in a Dirac material splits to form a pair of Weyl points with twofold degeneracy. Although Dirac and Weyl fermions in high energy physics are restricted by Poincaré invariance, their quasiparticle analogs in condensed matter physics are only constrained by the 230 space groups in three-dimensional lattices [12, 13]. Recently, it has been proposed that time-reversal and crystal symmetries in solids can protect exotic three-, six- or eightfold degenerate fermions not allowed in high energy physics [14]. One example is triply degenerate nexus fermions that can be considered as an intermediate state between Dirac and Weyl states [15-19]. Moreover, multifold chiral fermions were discovered in topological chiral crystals in the RhSi family, with long surface Fermi arcs stretching across the whole Brillouin zone (BZ) [20-24].

In this ever-evolving field, materials with the very common pyrite crystal structure have been proposed to show exotic properties. For example, Os$X_2$ ($X$ = Se, Te) was predicted to be topologically nontrivial with a band inversion near Γ due to strong spin-orbit coupling (SOC) interactions from Os [25]. Moreover, calculations show that the surface states of Os$X_2$ have both in-plane and out-of-plane spin components, which is different from spin textures observed in conventional topological insulators [25]. In addition, *ab initio* calculations predict that ferromagnetic CrO$_2$ with the pyrite structure can host type-II Weyl fermions, which would violate Lorentz invariance in vacuum [26]. In the presence of SOC, triply degenerate bands in CrO$_2$ split due to the mirror reflection symmetry breaking, resulting in Weyl fermions [26]. Recently, two photoemission studies explored another member of this family, PdSb$_2$ [27, 28], which is predicted to host sixfold degenerate band crossings at the corner of the BZ [14]. While these two works demonstrate bands near the multifold degeneracy, neither resolves all six bands due to the limitation of the total energy resolution. Therefore, decisive evidence for the sixfold degenerate fermion in PdSb$_2$ is still lacking. In this paper we use ultra-high resolution angle-resolved photoemission spectroscopy (ARPES) to unambiguously resolve all the degenerate bands at the R point. Supported by density functional theory (DFT) calculations, we find three doubly degenerate quadratic bands at the R point, visualizing the sixfold crossing.

Single crystal PdSb$_2$ samples were grown by the flux method [29]. The structure of as-grown crystals was examined through powder x-ray diffraction (XRD) using a PANalytical Empyrean x-ray diffractometer by crushing single crystals into fine powder. ARPES measurements were carried out at the beamline 5-2 of SSRL with circularly polarized light in the range of 25-48 eV.

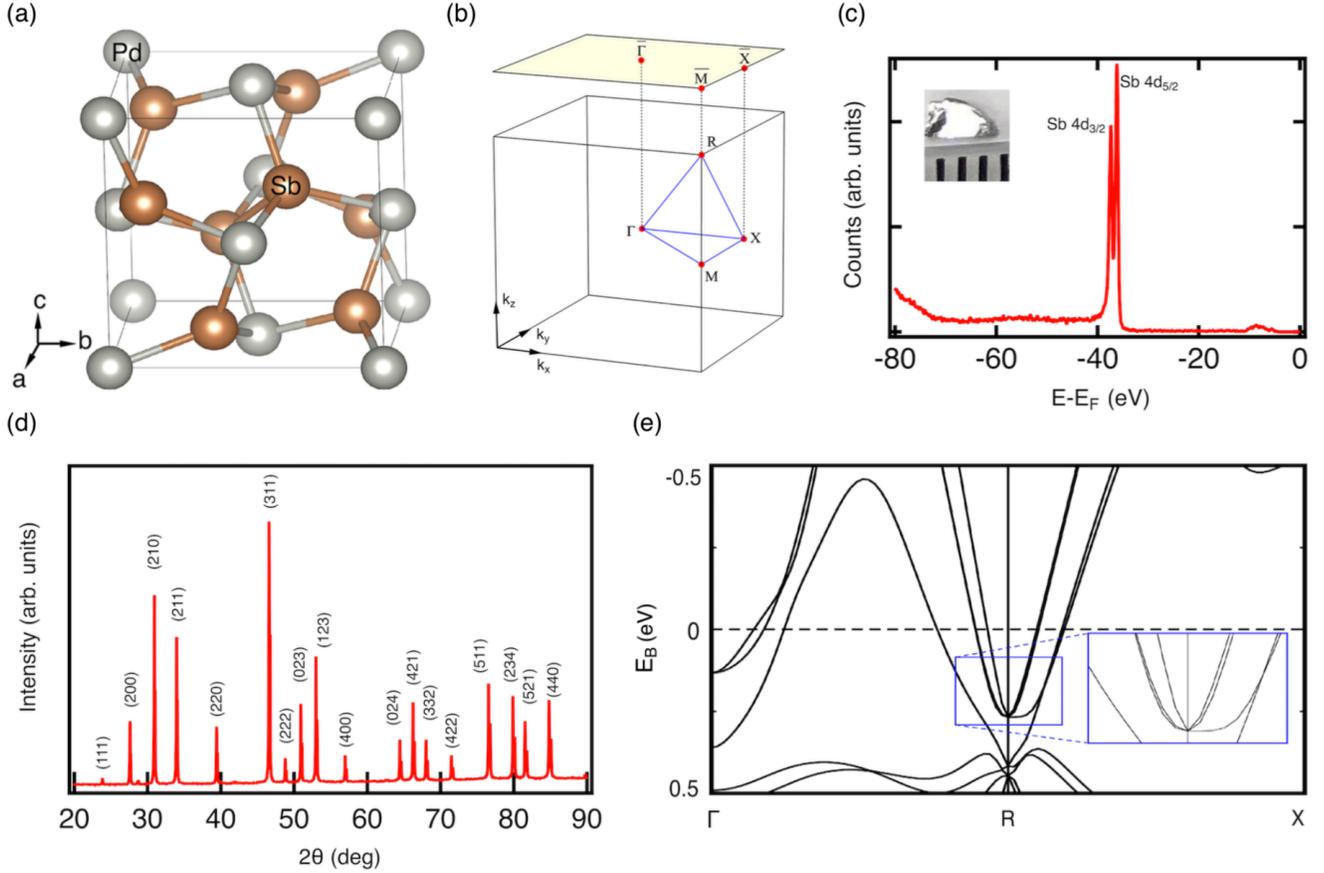

FIG. 1. Structure characterization of PdSb$_2$ single crystal and its band structure. (a) Crystal structure of PdSb$_2$. (b) Bulk and surface Brillouin zones of PdSb$_2$ with high symmetry points indicated. (c) Photoemission core level spectrum confirming the high quality of PdSb$_2$ crystals. Inset shows the surface of a typical PdSb$_2$ single crystal placed beside a millimeter scale. (d) Room temperature x-ray diffraction pattern of PdSb$_2$ showing the crystal structure. All diffraction peaks can be indexed under the pyrite type cubic structure with the space group #205. (e) Band structure calculations with spin-orbit coupling showing the sixfold degeneracy at the R point. The band structure near the R point in the blue box is enlarged to show the sixfold degeneracy.

Additional ARPES photon energy dependence measurements were performed at the Bloch beamline of the MAX IV Laboratory using linear horizontal polarized incident light in the range of 44-82 eV. The sample temperature was kept between 13 and 17 K at SSRL and 22-24 K at Bloch. Samples were cleaved *in situ* under a pressure lower than 5×10$^{-11}$ Torr, producing shiny surfaces. Photoelectrons were collected using a ScientaOmicron DA30 hemispherical analyzer, which resulted in a total energy resolution of less than 10 meV at SSRL and 28 meV at Bloch.

We performed first-principles calculations based on DFT. The structural and electronic properties of bulk PdSb$_2$ were all carried out using the Vienna *Ab Initio* Simulation Package (VASP) [30-33]. Ion cores were modeled with projector augmented wave pseudopotentials [34]. The valence 4$p$ and 4$d$ states of palladium and the 4$d$, 5$s$, and 5$p$ states of antimony were treated explicitly. A plane-wave basis energy cutoff of 450 eV and a Gaussian smearing of 0.2 eV were found to yield converged total energy and forces. We used the Perdew-Burke-Ernzerhof (PBE) exchange-correlation functional in our calculations [35]. Since the underlying symmetry of the space group protects the multifold degeneracy [14], PdSb$_2$ is cubic with the sixfold degeneracy occurring at the high symmetry R point, independent of *ab initio* methods used. Moreover, our *ab initio* calculations using the PBE functional do obtain the sixfold degeneracy as expected by symmetry. Therefore, DFT calculations using the PBE functional are sufficient to prove the sixfold degeneracy in PdSb$_2$. Furthermore, the SOC interactions were included in the self-consistent calculations. All atoms and the cell were relaxed to a force cutoff of $10^{-2}$ eV/Å. The $k$-point sampling was based on a $\Gamma$-centered grid and we used a (12×12×12) grid for all calculations. Maximally localized Wannier functions based on Pd-$d$ and Sb-$p$ orbitals were generated using Wannier90 code [36, 37]. The resulting electronic band structure reproduces that obtained from first-principles calculations at the vicinity of the Fermi energy.

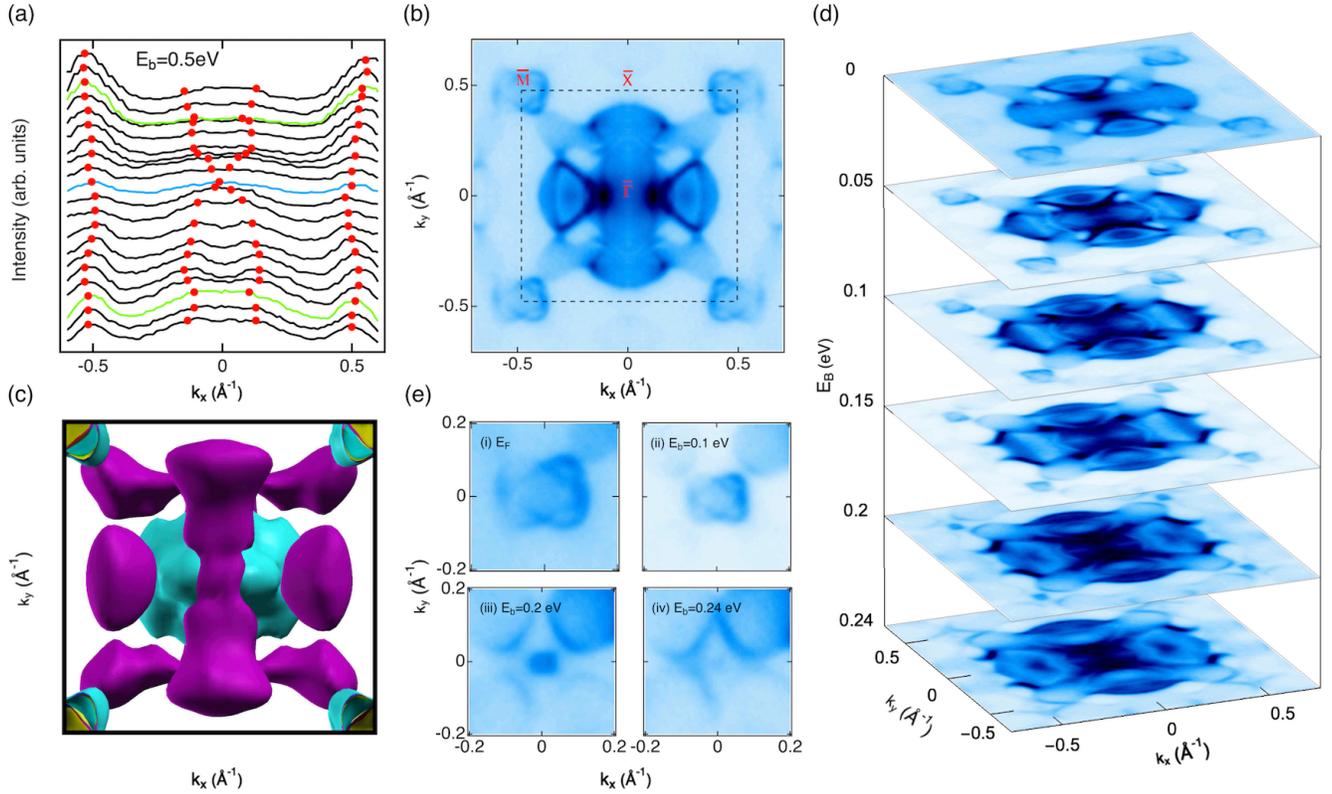

FIG. 2. Photon energy dependence and Fermi surface ARPES spectrum of PdSb$_2$. (a) Momentum distribution curves (MDCs) at various photon energies along the $\overline{X}$-$\overline{\Gamma}$-$\overline{X}$ high symmetry direction. Red dots represent peaks of MDCs extracted from the data. Blue and green MDCs correspond to $\Gamma$ and X points, respectively. (b) Symmetrized ARPES spectrum of PdSb$_2$ taken with 48 eV incident light. Black dotted box indicates the surface BZ marked with high symmetry points. (c) Calculated Fermi surface. (d) Symmetrized constant energy surfaces at photon energy of 48 eV at different binding energies. (e) Zoom-in plot of (d) around the R point demonstrating the evolution of the R point with various binding energies.

PdSb$_2$ crystal has the pyrite structure with the space group Pa$\overline{3}$ (#205). As a common crystal structure found in nature, pyrite type crystals are cubic and possess the compositional form AB$_2$. Pd atoms occupy the face-centered-cubic sites and have coordination number 6 (Fig. 1(a)). Each Sb atom has four nearest neighbors, three Pd atoms and one Sb atom. A photoemission core level spectrum demonstrates the characteristic peaks from d orbitals of Sb (Fig. 1(c)). Additionally, XRD suggests that samples have space group #205 with a lattice parameter equal to 6.464 Å (Fig. 1(d)), which is in good agreement with the literature [29]. In comparison with the experimental value of bulk PdSb$_2$, the theoretical result is slightly overestimated by ~1.6% (6.570 Å). To understand the electronic properties of PdSb$_2$, we calculated its bulk electronic band structure with SOC around the R point (Fig. 1(e)). There are three bands crossing at the time-reversal invariant R point, consistent with previous studies [14, 27-29]. In this case, nonsymmorphic symmetries are crucial to protect the degeneracy [14]. Moreover, the combination of time-reversal and inversion symmetries in PdSb$_2$ requires all bands to be doubly degenerate, which doubles the three-fold degeneracy at the R point to a sixfold degeneracy [14].

To experimentally probe a multifold degeneracy, we used vacuum ultraviolet ARPES to study the (001) surface. Aiming to precisely locate the photon energy corresponding to the R point, we first did a photon energy dependence in the $k_x$-$k_z$ plane with incident photon energies from 44 to 82 eV. Fig. 2(a) demonstrates corresponding momentum distribution curves (MDCs) along the $\overline{X}$-$\overline{\Gamma}$-$\overline{X}$ high symmetry direction, and the red points are determined from MDCs. We can clearly see spectral evolution at both the $\Gamma$ and X points indicating their bulk nature. Based on this photon energy dependence, we identified photon energy corresponding to the R point as 48 or 25 eV.

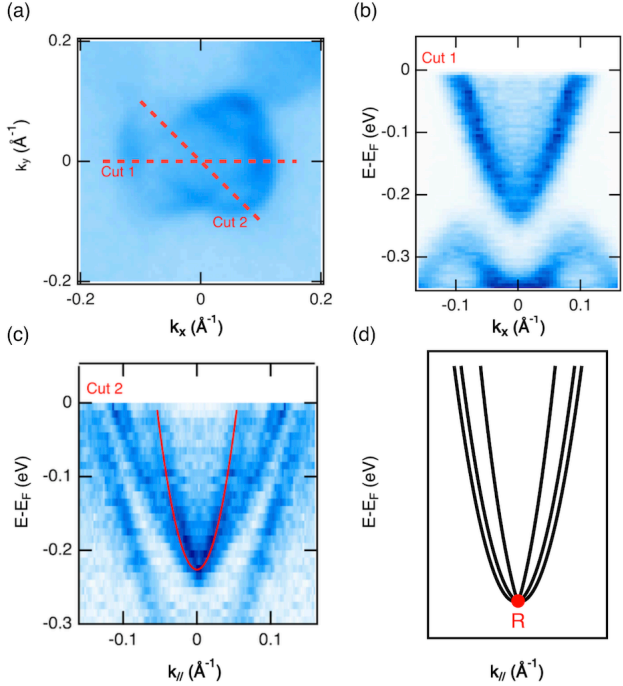

FIG. 3. Observation of sixfold degeneracy and quadratic dispersion of the R point in PdSb2. (a) Fermi surface spectrum focusing on the R point. Dotted lines define the k-space cut directions for Cuts 1 and 2. (b) Symmetrized dispersion map plot measured along the Cut 1 direction. (c) Symmetrized dispersion map plot measured along the Cut 2 direction. Red curve is the quadratic fit of the inner band based on the peaks of the corresponding momentum distribution curves. (d) Electronic band structure calculations along the same direction as (c) showing the sixfold degeneracy at the R point. Only the three doubly degenerate bands are shown and they are fitted with quadratic dispersion based on DFT calculations. The red dot represents the R point. All measurements were taken with 25 eV incident light except for (a) with 48 eV incident light.

Therefore, the Fermi surface map measured with 48 eV incident photons in Fig. 2(b) corresponds to the R-X-M plane in the bulk BZ. Furthermore, this Fermi surface map possesses the expected square surface BZ (Figs. 1(b) & 2(b)). Our experimental data match well with DFT calculations, with the $\overline{\Gamma}$ pocket in the center and a flower-shaped pattern at $\overline{M}$ (Figs. 2(b) & (c)). The experimentally measured Fermi surface also demonstrates the expected mirror symmetry and inversion symmetry of the space group. The electronic structure can be further tracked with increasing binding energy (Fig. 2(d)). Focusing on the well isolated bands at the $\overline{M}$ point, we observe their evolution (Fig. 2(e)). The pockets become smaller as binding energy is increased, indicating an electron-like dispersion. They finally shrink to a single point at binding energy 0.24 eV. This suggests that there may be a multifold fermion at a binding energy of 0.24 eV for PdSb2.

Fig. 3 is the main result of this article, demonstrating the sixfold degeneracy and quadratic dispersion at the R point. Due to the small splitting of the bands near the R point (Fig. 1(e)), we lowered the photon energy to 25 eV, corresponding to a full lattice vector translation in $k_z$. Fig. 3(b) shows the symmetrized cut through the R point along the Cut 1 direction noted in Fig. 3(a). Along this direction, all bands are very close to each other, which can also be seen from Fig. 3(a). It is thus hard to resolve them experimentally. Therefore, we measured the dispersion along the Cut 2 direction where bands are more spread out to better resolve the band splitting, as shown by Fig 3(c). Because of the low photon energy and high analyzer resolution during the measurement, the total energy resolution is lower than 3.8 meV for Fig. 3(c). With this extremely high energy resolution, we can resolve the three doubly degenerate bands crossing at the R point. The inner branch has one doubly degenerate band whereas the outer branch has two doubly degenerate bands, which can be seen near the Fermi level. Moreover, corresponding MDCs (Supplemental Material, Fig. S1 (b) [38]) suggest that these bands are truly degenerate, since they cross at the same point, which is consistent with theoretical predictions [14]. The extracted peaks from MDCs also confirm the three doubly degenerate bands. In addition, we calculated the electronic band structure along the same direction at the R point in Fig. 3(d). There are three doubly degenerate bands at the R Point. Furthermore, the two outer bands are closer to each other than the inner one, in agreement with experimental data (Fig. 3(c)). Therefore, we demonstrate that there are six bands crossing at the R point when spin-degeneracy is included. Moreover, the crossing is at a binding energy of 0.24 eV, consistent with Fig. 2(e) and calculations.

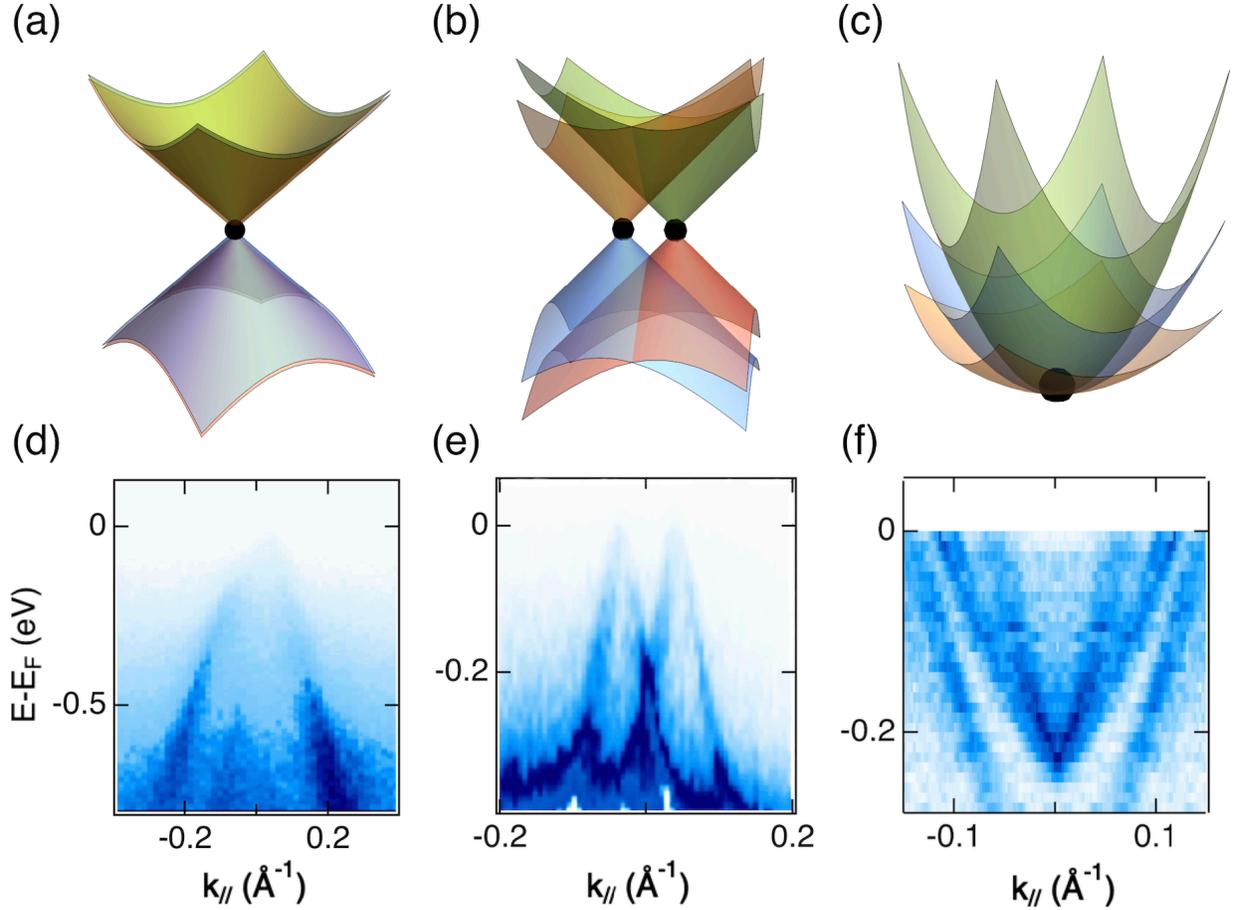

FIG. 4. Schematic representation of band structures for (a) Dirac fermion, (b) Weyl fermion and (c) sixfold degenerate fermion. Black dot represents the degenerate point. (d-f) Corresponding experimental ARPES dispersion maps showing (d) a Dirac node in $Na_3Bi$ [2], (e) two Weyl nodes in TaAs [9] and (f) a sixfold degenerate fermion in $PdSb_2$.

Due to both inversion and time-reversal symmetries in $PdSb_2$, it is shown that the lowest order of the dispersion near the multifold degeneracy has to be quadratic [14]. To show this quadratic dispersion around the R point, we fitted the inner band with only a constant term and a quadratic term in Fig. 3(c). The fit matches well with the data, which agrees with previous theoretical calculations [14]. Thus, we demonstrate that the bands around the R point have quadratic dispersion. The fitting also suggests that the crossing point at R is at about 0.23 eV below the Fermi level, in agreement with experimental data.

The novelty of this work is exemplified by comparing the sixfold fermion in $PdSb_2$ to previous topological fermions confirmed by ARPES measurements (Fig. 4). Dirac semimetals host three-dimensional (3D) massless electrons described by the Dirac equation where two doubly degenerate linear bands cross to form the Dirac point (Fig. 4(a)). One of the first Dirac semimetals that were experimentally confirmed by ARPES is $Na_3Bi$ [1-3] (Fig. 4(d)). The fourfold degenerate Dirac crossings have net Chern number zero. When either inversion symmetry or time-reversal symmetry is broken in a 3D material, Weyl fermions can form, which are doubly degenerate (Fig. 4(b)). Unlike Dirac fermions, Weyl fermions carry nonzero Chern number. They can be viewed as sources and sinks of the Berry curvature so Weyl nodes with opposite chirality must appear in pairs. TaAs is the first experimental realization of Weyl fermions in solid state physics [9-11], where two singly degenerate linear bands cross to form a Weyl fermion (Fig. 4(e)). While both Dirac and Weyl fermions have counterparts in high energy physics, sixfold fermions are not allowed by Poincaré symmetry but are possible in solid state systems. In $PdSb_2$, the sixfold degeneracy at the time-reversal invariant R point is protected by its nonsymmorphic symmetry [14]. Bands are quadratic near the R point and the net Chern number is zero, as indicated by Figs. 4(c) and 4(f). Moreover, $PdSb_2$ becomes superconducting under quasihydrostatic pressure [29]. Therefore, further study can be done to understand the interaction between bosons (Cooper pairs) and sixfold fermions in this system. In addition, it is shown that magnetic order can break a sixfold degeneracy to three- or twofold crossings [39]. For example, Fe-doped PtSb$_2$ with pyrite structure can host doubly degenerate magnetic Weyl nodes [39]. Furthermore, previous de Haas-van Alphen oscillations have probed nearly massless particles with nontrivial Berry phase in PdSb$_2$ with the magnetic field applied along the [111] direction [29]. Therefore, it is interesting to break time-reversal symmetry in PdSb$_2$ by an external magnetic field or doping to create three- or twofold Weyl nodes [29, 39]. Further scanning tunneling microscopy measurements can also be performed to study the properties of PdSb$_2$ under magnetic field [40, 41].

In conclusion, using high resolution ARPES supported by *ab initio* calculations, we clearly demonstrate a sixfold degeneracy at the R point in PdSb$_2$. Furthermore, we show the bands near the R point are quadratic. Our study indicates a sixfold fermion beyond the constraint of high energy physics and motivates further investigations of exotic fermions.

Work at Princeton University and Princeton-led synchrotron based ARPES measurements are supported by the U.S. Department of Energy under Grant No. DOE/BES DE-FG-02-05ER46200. Work at LSU is supported by the U.S. Department of Energy (DOE) under EPSCoR Grant No. DE-SC0012432 with additional support from the Louisiana Board of Regents. Part of this work was performed using supercomputing resources provided by Louisiana State University and the Center for Computational Innovations (CCI) at Rensselaer Polytechnic Institute. Use of the Stanford Synchrotron Radiation Lightsource, SLAC National Accelerator Laboratory, is supported by the U.S. DOE, Office of Science, Office of Basic Energy Sciences under Contract No. DE-AC02-76SF00515. The authors thank D. Lu and M. Hashimoto for beamtime support at SSRL Beamline 5-2. The authors thank the MAX IV Laboratory for access to the Bloch Beamline under proposal No. 20190502. The authors also thank B. Thiagarajan, C. Polley, H. Fedderwitz and J. Adell for beamtime support at Bloch. T.A.C. acknowledges the support of the NSF Graduate Research Fellowship Program (DGE-1656466).

* These authors contributed equally to this work.
† Corresponding authors: xiany@princeton.edu; mzhasan@princeton.edu

[1] S.-Y. Xu, C. Liu, S. K. Kushwaha, R. Sankar, J. W. Krizan, I. Belopolski, M. Neupane, G. Bian, N. Alidoust, T.-R. Chang, H.-T. Jeng, C.-Y. Huang, W.-F. Tsai, H. Lin, P. P. Shibayev, F.-C. Chou, R. J. Cava, and M. Z. Hasan, Science **347**, 294 (2015).

[2] S.-Y. Xu, C. Liu, I. Belopolski, S. K. Kushwaha, R. Sankar, J. W. Krizan, T.-R. Chang, C. M. Polley, J. Adell, T. Balasubramanian, K. Miyamoto, N. Alidoust, G. Bian, M. Neupane, H.-T. Jeng, C.-Y. Huang, W.-F. Tsai, T. Okuda, A. Bansil, F. C. Chou, R. J. Cava, H. Lin, and M. Z. Hasan, Phys. Rev. B **92**, 075115 (2015).

[3] Z. K. Liu, B. Zhou, Y. Zhang, Z. J. Wang, H. M. Weng, D. Prabhakaran, S.-K. Mo, Z. X. Shen, Z. Fang, X. Dai, Z. Hussain, and Y. L. Chen, Science **343**, 864 (2014).

[4] M. Neupane, S.-Y. Xu, R. Sankar, N. Alidoust, G. Bian, C. Liu, I. Belopolski, T.-R. Chang, H.-T. Jeng, H. Lin, A. Bansil, F. C. Chou, and M. Z. Hasan, Nat. Commun. **5**, 3786 (2014).

[5] Z. K. Liu, J. Jiang, B. Zhou, Z. J. Wang, Y. Zhang, H. M. Weng, D. Prabhakaran, S.-K. Mo, H. Peng, P. Dudin, T. Kim, M. Hoesch, Z. Fang, X. Dai, Z. X. Shen, D. L. Feng, Z. Hussain, and Y. L. Chen, Nat. Mater. **13**, 677 (2014).

[6] S.-M. Huang, S.-Y. Xu, I. Belopolski, C.-C. Lee, G. Chang, B. Wang, N. Alidoust, G. Bian, M. Neupane, C. Zhang, S. Jia, A. Bansil, H. Lin, and M. Z. Hasan, Nat. Commun. **6**, 7373 (2015).

[7] H. Weng, C. Fang, Z. Fang, B. A. Bernevig, and X. Dai, Phys. Rev. X **5**, 011029 (2015).

[8] S.-Y. Xu, N. Alidoust, I. Belopolski, Z. Yuan, G. Bian, T.-R. Chang, H. Zheng, V. N Strocov, D. S. Sanchez, G. Chang, C. Zhang, D. Mou, Y. Wu, L. Huang, C.-C. Lee, S.-M. Huang, B. Wang, A. Bansil, H.-T. Jeng, T. Neupert, A. Kaminski, H. Lin, S. Jia, and M. Z. Hasan, Nat. Phys. **11**, 748 (2015).

[9] S.-Y. Xu, I. Belopolski, N. Alidoust, M. Neupane, G. Bian, C. Zhang, R. Sankar, G. Chang, Z. Yuan, C.-C. Lee, S.-M. Huang, H. Zheng, J. Ma, D. S. Sanchez, B. Wang, A. Bansil, F. Chou, P. P. Shibayev, H. Lin, S. Jia, and M. Z. Hasan, Science **349**, 613 (2015).

[10] B. Q. Lv, H. M. Weng, B. B. Fu, X. P. Wang, H. Miao, J. Ma, P. Richard, X. C. Huang, L. X. Zhao, G. F. Chen, Z. Fang, X. Dai, T. Qian, and H. Ding, Phys. Rev. X **5**, 031013 (2015).

[11] L. X. Yang, Z. K. Liu, Y. Sun, H. Peng, H. F. Yang, T. Zhang, B. Zhou, Y. Zhang, Y. F. Guo, M. Rahn, D. Prabhakaran, Z. Hussain, S.-K. Mo, C. Felser, B. Yan, and Y. L. Chen, Nat. Phys. **11**, 728 (2015).

[12] F. Tang, H. C. Po, A. Vishwanath, and X. Wan, Nature (London) **566,** 486 (2019).

[13] T. Zhang, Y. Jiang, Z. Song, H. Huang, Y. He, Z. Fang, H. Weng, and C. Fang, Nature (London) **566,** 475 (2019).

[14] B. Bradlyn, J. Cano, Z. Wang, M. G. Vergniory, C. Felser, R. J. Cava, and B. A. Bernevig, Science **353**, aaf5037 (2016).

[15] G. Chang, S.-Y. Xu, S.-M. Huang, D. S. Sanchez, C.-H. Hsu, G. Bian, Z.-M. Yu, I. Belopolski, N. Alidoust, H. Zheng, T.-R. Chang, H.-T. Jeng, S. A. Yang, T. Neupert, H. Lin, and M. Z. Hasan, Sci. Rep. **7**, 1688 (2017).

[16] H. Weng, C. Fang, Z. Fang, and X. Dai, Phys. Rev. B **93**, 241202(R) (2016).

[17] Z. Zhu, G. W. Winkler, Q. S. Wu, J. Li, and A. A. Soluyanov, Phys. Rev. X **6**, 031003 (2016).

[18] B. Q. Lv, Z.-L. Feng, Q.-N. Xu, X. Gao, J.-Z. Ma, L.-Y. Kong, P. Richard, Y.-B. Huang, V. N. Strocov, C. Fang, H.-M. Weng, Y.-G. Shi, T. Qian, and H. Ding, Nature (London) **546**, 627 (2017).

[19] J.-Z. Ma, J.-B. He, Y.-F. Xu, B. Q. Lv, D. Chen, W.-L. Zhu, S. Zhang, L.-Y. Kong, X. Gao, L.-Y. Rong, Y.-B. Huang, P. Richard, C.-Y. Xi, E. S. Choi, Y. Shao, Y.-L. Wang, H.-J. Gao, X. Dai, C. Fang, H.-M. Weng, G.-F. Chen, T. Qian, and H. Ding, Nat. Phys. **14**, 349 (2018).

[20] G. Chang, S.-Y. Xu, B. J. Wieder, D. S. Sanchez, S.-M. Huang, I. Belopolski, T.-R. Chang, S. Zhang, A. Bansil, H. Lin, and M. Z. Hasan, Phys. Rev. Lett. **119**, 206401 (2017).

[21] D. S. Sanchez, I. Belopolski, T. A. Cochran, X. Xu, J.-X. Yin, G. Chang, W. Xie, K. Manna, V. Süß, C.-Y. Huang, N. Alidoust, D. Multer, S. S. Zhang, N. Shumiya, X. Wang, G.-Q. Wang, T.-R. Chang, C. Felser, S.-Y. Xu, S. Jia, H. Lin, and M. Z. Hasan, Nature (London) **567**, 500 (2019).

[22] Z.-C. Rao, H. Li, T.-T. Zhang, S.-J. Tian, C.-H. Li, B.-B. Fu, C.-Y. Tang, L. Wang, Z.-L. Li, W.-H. Fan, J.-J. Li, Y.-B. Huang, Z.-H. Liu, Y.-W. Long, C. Fang, H.-M. Weng, Y.-G. Shi, H.-C. Lei, Y.-J. Sun, T. Qian, and H. Ding, Nature (London) **567**, 496 (2019).

[23] D. Takane, Z. Wang, S. Souma, K. Nakayama, T. Nakamura, H. Oinuma, Y. Nakata, H. Iwasawa, C. Cacho, T. Kim, K. Horiba, H. Kumigashira, T. Takahashi, Y. Ando, and T. Sato, Phys. Rev. Lett. **122**, 076402 (2019).

[24] N. B. M. Schröter, D. Pei, M. G. Vergniory, Y. Sun, K. Manna, F. de Juan, J. A. Krieger, V. Süss, M. Schmidt, P. Dudin, B. Bradlyn, T. K. Kim, T. Schmitt, C. Cacho, C. Felser, V. N. Strocov, and Y. Chen, Nat. Phys. **15**, 759 (2019).

[25] Y. Yin, M. S. Fuhrer, and N. V. Medhekar, npj Quantum Mater. **4,** 47 (2019).

[26] R. Wang, Y. J. Jin, J. Z. Zhao, Z. J. Chen, Y. J. Zhao, and H. Xu, Phys. Rev. B **97**, 195157 (2018).

[27] N. Kumar, M. Yao, J. Nayak, M. G. Vergniory, J. Bannies, Z. Wang, N. B. M. Schröter, V. N. Strocov, L. Müchler, W. Shi, E. D. L. Rienks, J. L. Mañes, C. Shekhar, S. S. P. Parkin, J. Fink, G. H. Fecher, Y. Sun, B. A. Bernevig, and C. Felser, Adv. Mater. **32**, 1906046 (2020).

[28] Z. P. Sun, C. Q. Hua, X. L. Liu, Z. T. Liu, M. Ye, S. Qiao, Z. H. Liu, J. S. Liu, Y. F. Guo, Y. H. Lu, and D. W. Shen, Phys. Rev. B **101**, 155114 (2020).

[29] R. Chapai, Y. Jia, W. A. Shelton, R. Nepal, M. Saghayezhian, J. F. DiTusa, E. W. Plummer, C. Jin, and R. Jin, Phys. Rev. B **99**, 161110(R) (2019).

[30] G. Kresse and J. Hafner, Phys. Rev. B **47**, 558(R) (1993).

[31] G. Kresse and J. Hafner, Phys. Rev. B **49**, 14251 (1994).

[32] G. Kresse and J. Furthmüller, Phys. Rev. B **54**, 11169 (1996).

[33] G. Kresse and J. Furthmüller, Comput. Mater. Sci. **6**, 15 (1996).

[34] P. E. Blöchl, Phys. Rev. B **50**, 17953 (1994).

[35] G. Kresse and D. Joubert, Phys. Rev. B **59**, 1758 (1999).

[36] A. A. Mostofi, J. R. Yates, G. Pizzi, Y.-S. Lee, I. Souza, D. Vanderbilt, and N. Marzari, Comput. Phys. Commun. **185**, 2309 (2014).

[37] G. Pizzi, V. Vitale, R. Arita, S. Blügel, F. Freimuth, G. Géranton, M. Gibertini, D. Gresch, C. Johnson, T. Koretsune, *et al.*, J. Phys.: Condens. Matter **32**, 165902 (2020).

[38] See Supplemental Material at http://link.aps.org/supplemental/ for the original spectrum of sixfold degenerate bands at the R point and their corresponding MDCs.

[39] M. G. Vergniory, L. Elcoro, F. Orlandi, B. Balke, Y.-H. Chan, J. Nuss, A. P. Schnyder, and L. M. Schoop, Eur. Phys. J. B **91**, 213 (2018).

[40] J.-X. Yin, S. S. Zhang, H. Li, K. Jiang, G. Chang, B. Zhang, B. Lian, C. Xiang, I. Belopolski, H. Zheng, T. A. Cochran, S.-Y. Xu, G. Bian, K. Liu, T.-R. Chang, H. Lin, Z.-Y. Lu, Z. Wang, S. Jia, W. Wang, and M. Zahid Hasan, Nature (London) **562**, 91 (2018).

[41] J.-X. Yin, S. S. Zhang, G. Chang, Q. Wang, S. S. Tsirkin, Z. Guguchia, B. Lian, H. Zhou, K. Jiang, I. Belopolski, N. Shumiya, D. Multer, M. Litskevich, T. A. Cochran, H. Lin, Z. Wang, T. Neupert, S. Jia, H. Lei, and M. Zahid Hasan, Nat. Phys. **15**, 443 (2019).